# Interpolable Formulas in Equilibrium Logic and Answer Set Programming

**Dov Gabbay**                                                DOV.GABBAY@KCL.AC.UK
*Bar Ilan University Israel, King's College London and*
*University of Luxembourg.*
**David Pearce**                                              DAVID.PEARCE@UPM.ES
*AI Dept, Universidad Politécnica de Madrid, Spain.*
**Agustín Valverde**                                          A_VALVERDE@CTIMA.UMA.ES
*Dept of Applied Mathematics, Universidad de Málaga, Spain.*

## Abstract

Interpolation is an important property of classical and many non-classical logics that has been shown to have interesting applications in computer science and AI. Here we study the Interpolation Property for the the non-monotonic system of *equilibrium logic*, establishing weaker or stronger forms of interpolation depending on the precise interpretation of the inference relation. These results also yield a form of interpolation for ground logic programs under the answer sets semantics. For disjunctive logic programs we also study the property of *uniform* interpolation that is closely related to the concept of variable forgetting. The first-order version of equilibrium logic has analogous Interpolation properties whenever the collection of equilibrium models is (first-order) definable. Since this is the case for so-called *safe* programs and theories, it applies to the usual situations that arise in practical answer set programming.

## 1. Introduction

The interpolation property is an important and much discussed topic in logical systems, both classical and non-classical (Gabbay & Maksimova, 2005). Its importance in computer science is also becoming recognised nowadays. The interpolation property has been applied in various areas of computer science, for example in software specification (Diaconescu, Goguen, & Stefaneas, 1993; Bicarregui, Dimitrakos, Gabbay, & Maibaum, 2001), in the construction of formal ontologies (Kontchakov, Wolter, & Zakharyaschev, 2008) and in model checking and related subareas (McMillan, 2005). In the first two areas interpolation is important as a metatheoretical property, in particular it may provide a basis for the modular composition and decomposition of theories; for instance, for Kontchakov et al. (2008) it plays a key role in the study of the modular decomposition of ontologies. In other cases, interpolants themselves play a role as special formulas applied in automated deduction (McMillan, 2005).

To date interpolation has received much less attention in systems of non-monotonic reasoning and logic programming, despite their importance in AI and computer science. In this note we study the interpolation property for the system of nonmonotonic reasoning known as *equilibrium logic* (Pearce, 2006). Since this in turn can be regarded as a logical foundation for stable model reasoning and answer set programming (ASP), our results transfer immediately to the sphere of ASP. We shall focus here mainly on interpolation as a metatheoretical property and our primary interest is in establishing this property for certain cases of interest. Although in Section 8 we consider a case where an interpolant (actually a uniform interpolant) is explicitly constructed, we are mainly concerned here with pure existence theorems. Discussion of complexity issues as well as possible applications of the interpolation property in ASP are left to future work. However, it seems likely that, as in the case of studies involving formal ontologies (Konev, Walther, & Wolter, 2009), interpolation may be a useful property for applications of ASP in knowledge representation. In a previous paper (Pearce & Valverde, 2012), the interpolation and Beth properties of the underlying, monotonic logic of ASP were used to characterise strong kinds of intertheory relations. To capture weaker kinds of intertheory relations it may be important to be able to rely on interpolation holding in the extended, non-monotonic logic. We plan to explore this avenue in the future.

To introduce the property of interpolation, let us start with some notation and terminology. Let us assume the syntax of first-order logic with formulas denoted by lower case Greek letters and predicates by lower case Latin letters.

Let $\vdash$ be a monotonic inference relation and suppose that $\alpha \vdash \beta$. An *interpolant for* $(\alpha, \beta)$ is a formula $\gamma$ such that

$$\alpha \vdash \gamma \;\; \& \;\; \gamma \vdash \beta \tag{1}$$

where $\gamma$ contains only predicate and constant symbols that belong to both $\alpha$ and $\beta$. A logic $L$ with inference relation $\vdash_L$ is said to have the *interpolation property* if an interpolant exists for every pair of formulas $(\alpha, \beta)$ such that $\alpha \vdash_L \beta$. As is well-known, classical logic as well as many non-classical logics possess interpolation.

Suppose now we deal with a non-monotonic logical system with an inference relation $\vdash\!\!\sim$. To express the idea that a formula is an interpolant it will not generally suffice simply to replace $\vdash$ by $\vdash\!\!\sim$ in (1). One problem is that, since $\vdash\!\!\sim$ is non-monotonic, it is in general not transitive. Instead, following the idea of Gabbay and Maksimova (2005), we can modify condition (1) and proceed in a two-stage fashion. We make use of the fact that non-monotonic consequence can be defined in terms of minimal models in some monotonic logical system, say that the consequence relation $\vdash\!\!\sim$ is appropriately captured by means of minimal models in a logic $L$ with consequence relation $\models_L$. Now suppose that $\alpha \vdash\!\!\sim \beta$. Then as an interpolant for $(\alpha, \beta)$ we look for a formula $\gamma$ such that

$$\alpha \vdash\!\!\sim \gamma \;\; \& \;\; \gamma \models_L \beta \tag{2}$$

where all predicate and constant symbols of $\gamma$ occur in both $\alpha$ and $\beta$. Since $\vdash\!\!\sim$ is to be defined via a subclass of minimal $L$-models, we already assume that $\models_L \subseteq \vdash\!\!\sim$. Moreover we should assume too that $L$ is a well-behaved sublogic in the sense that $L$-equivalent formulas have the same $\vdash\!\!\sim$-consequences and that the $L$-consequences of $\vdash\!\!\sim$-consequences are themselves $\vdash\!\!\sim$-consequences (so e.g. from (2) we can derive $\alpha \vdash\!\!\sim \beta$). In non-monotonic reasoning these last two properties are known as left and right absorption, respectively. Given these conditions, it follows at once from (2) that any formula in the language of $\gamma$ that is $L$-equivalent to $\gamma$ will also be an interpolant for $(\alpha, \beta)$. Likewise if $\gamma$ is an interpolant for $(\alpha, \beta)$ and $\beta \models_L \delta$ then $\alpha \vdash\!\!\sim \delta$ and $\gamma$ is an interpolant for $(\alpha, \delta)$.

Now, to find an interpolant for $(\alpha, \beta)$ satisfying (2), or to prove that one always exists, we can proceed as follows. We look for an $L$-formula $\alpha'$ say, that precisely $L$-defines the minimal models of $\alpha$. Since $\alpha \mathrel{|\!\!\!\sim} \beta$ it follows that $\alpha' \models_L \beta$. Now, if $L$ has the interpolation property as defined earlier, we apply this theorem to obtain or infer the existence of an $L$-interpolant $\gamma$ in the sense of (1) for $(\alpha', \beta)$. Hence (2) follows.

Notice that this two-stage procedure relies on three key features: (i) that we can identify a suitable monotonic sublogic $L$ for $\mathrel{|\!\!\!\sim}$, (ii) that a formula's minimal models are $L$-definable, and (iii) that $L$ has the interpolation property. These conditions are *prima facie* independent. As we shall see, we may have (i) and (iii) but lack (ii). The situation with respect to equilibrium logic is as follows. In the propositional case all three conditions are met, so we can establish the interpolation property in the general case. The situation for quantified equilibrium logic is more complicated. In the general case, we lack condition (ii). More precisely, we have an appropriate monotonic sublogic $L$ and this logic has the interpolation property, but the equilibrium models of a formula need not be first-order definable in $L$. So the procedure outlined does not yield interpolants in all cases. However some recent results on a generalised concept of (first-order) stable model imply that there are interesting classes of interpolable formulas: we shall consider in more detail one such class, that of *safe* formulas. In particular, if $\alpha$ is a safe formula and $\alpha \mathrel{|\!\!\!\sim} \beta$, then there exists an interpolant $\gamma$ such that (2) holds. Other classes of interpolable formulas are so-called *tight* formulas, and formulas possessing a finite, complete set of what are called *loops*.

Safety, tightness and loop formulas have been studied at some length in answer set programming (ASP). The implications of these results for ASP can be summarised as follows. In the case of (finite) ground programs the interpolation property holds. In the first-order or non-ground case, interpolation holds for finite, safe programs without function symbols, and hence practically for all finite programs currently admitted by answer set solvers. Moreover, since safety is now defined for arbitrary formulas in a function-free language, the class of safe formulas in this sense goes beyond the class of expressions normally admitted in ASP, even if auxiliary concepts like weight constraints and aggregates are included.

## 2. Logical Preliminaries

We work with standard propositional and predicate languages, where the latter may in the general case contain constant and function symbols. Propositional languages are based on a set $V$ of propositional variables, and formulas are built-up in the usual way using the logical constants $\wedge$, $\vee$, $\rightarrow$, $\neg$, standing respectively for conjunction, disjunction, implication and negation. If $\varphi$ is a propositional formula, we denote by $V(\varphi)$ the set of propositional variables appearing in $\varphi$.

A *first-order language* $\mathcal{L} = \langle C, F, P \rangle$ consists of a set of constants $C$, function symbols $F$ and predicate symbols $P$; each function symbol $f$ in $F$ and predicate symbol $p \in P$ has an assigned arity. Moreover, we assume a fixed countably infinite set of variables, the symbols '$\rightarrow$', '$\vee$', '$\wedge$', '$\neg$', '$\exists$', '$\forall$', and auxiliary parentheses '(', ')'. *Atoms*, *terms* and *formulas* are constructed as usual; *closed* formulas, or *sentences*, are those where each variable is bound by some quantifier. If $\varphi$ is a (first-order) formula, $\mathcal{L}(\varphi)$ denotes the language associated with $\varphi$, i.e. the set of constants, function and predicate symbols occuring in it.

We make use of the following notation and terminology. Boldface $\mathbf{x}$ stands for a tuple of variables, $\mathbf{x} = (x_1, \ldots, x_n)$, while $\varphi(\mathbf{x}) = \varphi(x_1, \ldots, x_n)$ is a formula whose free variables are $x_1, \ldots, x_n$, and $\forall \mathbf{x} = \forall x_1 \ldots \forall x_n$. If $t_i$ are terms, then $\mathbf{t} = (t_1, \ldots, t_n)$ denotes a *vector* of terms. A *theory* $\Pi$ is a set of sentences. Variable-free terms, atoms, formulas, or theories are also called *ground*.

As usual the symbols $\vdash$ and $\models$, possibly with subscripts, are used to denote logical inference and consequence relations, respectively. A logic $L$ is said to be *monotonic* if its inference relation $\vdash_L$ satisfies the monotonicity property:

$$\Pi \vdash_L \varphi \ \& \ \Pi \subseteq \Pi' \ \Rightarrow \ \Pi' \vdash_L \varphi$$

To distinguish non-monotonic from monotonic inference relations, we use $\vdash\!\!\!\sim$ to symbolise the former. In most cases a non-monotonic logic can be understood in terms of an inference relation that extends a suitable monotonic logic. When this extension is well-behaved we say that the monotonic logic forms a *deductive base*[1] (Pearce, 2006) for it. This can be made precise as follows.

**Definition 1** *Let $\vdash\!\!\!\sim$ be any nonmonotonic inference relation. We say that a logic $L$ with monotonic inference relation $\vdash_L$ is a deductive base for $\vdash\!\!\!\sim$ iff (i) $\vdash_L \subseteq \vdash\!\!\!\sim$; (ii) If $\Pi_1 \equiv_L \Pi_2$ then $\Pi_1 \approx \Pi_2$; (iii) If $\Pi \vdash\!\!\!\sim \varphi$ and $\varphi \vdash_L \psi$, then $\Pi \vdash\!\!\!\sim \psi$.*

Here $\equiv_L$ denotes ordinary logical equivalence in $L$, while $\approx$ denotes non-monotonic equivalence, i.e. $\Pi_1 \approx \Pi_2$ means that $\Pi_1$ and $\Pi_2$ have the same non-monotonic consequences. Furthermore, we say that a deductive base is *strong* if it satisfies the additional condition:

$$\Pi_1 \not\equiv_L \Pi_2 \Rightarrow \quad \text{there exists } \Gamma \text{ such that } \Pi_1 \cup \Gamma \not\approx \Pi_2 \cup \Gamma.$$

In terms of nonmonotonic consequence operations, (ii) and (iii) correspond to conditions known as left absorption and right absorption respectively (Makinson, 1994).

## 2.1 Interpolation

We now turn to the interpolation property.

**Definition 2** *A logic $L$ with inference relation $\vdash_L$ is said to have the* interpolation property *if whenever*

$$\vdash_L \varphi \to \psi$$

*there exists a sentence $\xi$ (the* interpolant*) such that*

$$\vdash_L \varphi \to \xi \ \& \ \vdash_L \xi \to \psi$$

*where all predicate, function and constant symbols of $\xi$ are contained in both $\varphi$ and $\psi$, i.e. $\mathcal{L}(\xi) \subseteq \mathcal{L}(\varphi) \cap \mathcal{L}(\psi)$. In the case of propositional logic, the requirement is that $V(\xi) \subseteq V(\varphi) \cap V(\psi)$.*

As explained in the introduction, for non-monotonic logics we can consider two forms of interpolation, one weaker one stronger. The stronger form makes use of an underlying monotonic logic.

---

1. It is close to the concept of *fully absorbing inferential frame* used by Dietrich (1994).

**Definition 3** *Suppose that $\alpha \mathbin{\mid\!\sim} \beta$. A $(\mathbin{\mid\!\sim}, \vdash_L)$ interpolant for $(\alpha, \beta)$ is a formula $\gamma$ such that*

$$\alpha \mathbin{\mid\!\sim} \gamma \ \ \& \ \ \gamma \vdash_L \beta \tag{3}$$

*where $L$ is a deductive base for $\mathbin{\mid\!\sim}$ and $\gamma$ contains only predicate, function and constant symbols that belong to both $\alpha$ and $\beta$. A non-monotonic logic with inference relation $\mathbin{\mid\!\sim}$ is said to have the $(\mathbin{\mid\!\sim}, \vdash)$ interpolation property if for a suitable deductive base logic $L$ an $(\mathbin{\mid\!\sim}, \vdash_L)$ interpolant exists for every pair of formulas $(\alpha, \beta)$ such that $\alpha \mathbin{\mid\!\sim} \beta$.*

The requirement that $L$ form a deductive base ensures that some desirable properties of interpolation are met.

**Proposition 1** *Let $\gamma$ be a $(\mathbin{\mid\!\sim}, \vdash_L)$ interpolant for $(\alpha, \beta)$.*
    *(a) For any $\psi$ such that $\psi \equiv_L \gamma$, $\psi$ is a $(\mathbin{\mid\!\sim}, \vdash_L)$ interpolant for $(\alpha, \beta)$.*
    *(b) For any $\alpha'$ such that $\alpha \equiv_L \alpha'$, and any $\beta'$ such that $\beta \vdash_L \beta'$, $\gamma$ is a $(\mathbin{\mid\!\sim}, \vdash_L)$ interpolant for $(\alpha', \beta')$.*

The property of deductive base also guarantees that the $(\mathbin{\mid\!\sim}, \vdash_L)$ relation is transitive in the sense that if (3) holds for any $\alpha, \beta, \gamma$, then also $\alpha \mathbin{\mid\!\sim} \beta$. This last property will not necessarily hold for the second, weaker form of interpolation that we call $(\mathbin{\mid\!\sim}, \mathbin{\mid\!\sim})$ interpolation.

**Definition 4** *Suppose that $\alpha \mathbin{\mid\!\sim} \beta$. A $(\mathbin{\mid\!\sim}, \mathbin{\mid\!\sim})$ interpolant for $(\alpha, \beta)$ is a formula $\gamma$ such that*

$$\alpha \mathbin{\mid\!\sim} \gamma \ \ \& \ \ \gamma \mathbin{\mid\!\sim} \beta \tag{4}$$

*where $\gamma$ contains only predicate, function and constant symbols that belong to both $\alpha$ and $\beta$. In the case of propositional logic, the requirement is that $V(\xi) \subseteq V(\varphi) \cap V(\psi)$.*

Analogous to the previous case, we say that a non-monotonic logic with inference relation $\mathbin{\mid\!\sim}$ has the $(\mathbin{\mid\!\sim}, \mathbin{\mid\!\sim})$ *interpolation property* if a $(\mathbin{\mid\!\sim}, \mathbin{\mid\!\sim})$ interpolant exists for every pair of formulas $(\alpha, \beta)$ such that $\alpha \mathbin{\mid\!\sim} \beta$. Notice that $(\mathbin{\mid\!\sim}, \vdash)$ is the stronger form of interpolation because if a logic has $(\mathbin{\mid\!\sim}, \vdash)$ interpolation it must also have $(\mathbin{\mid\!\sim}, \mathbin{\mid\!\sim})$ interpolation, again as a consequence of the deductive base requirement (first clause).

Evidently the properties expressed in Proposition 1 are not directly applicable to the second form of interpolation that does not refer to any underlying base logic. Nevertheless an important feature of the interpolation properties we shall establish below is that we can formulate and prove analogous properties even for $(\mathbin{\mid\!\sim}, \mathbin{\mid\!\sim})$ interpolation.

We can also consider restricted variants of interpolation when the property holds for certain types of formulas, in other words, when there is an interpolant for $(\alpha, \beta)$ given $\alpha \mathbin{\mid\!\sim} \beta$ whenever $\alpha$ and $\beta$ belong to specific syntactic classes. In such cases we can refer to *interpolable* formulas. Later on we shall consider both kinds of restrictions, where $\alpha$ belongs to a specific class or alternatively when $\beta$ does.

## 2.2 Review of the Logic of Here-and-There

Equilibrium logic is based on the nonclassical logic of here-and-there, which we denote by **HT** in the propositional case. In the quantified or first-order case we denote the logic by **QHT**, with subscripts/superscripts to denote specific variants.

In both propositional and quantified cases the logic is based on the axioms and rules of intuitionistic logic and is captured by the usual Kripke semantics for intuitionistic logic (van Dalen, 1997). However the additional axioms of **HT** and **QHT** mean that we can use very simple kinds of Kripke structures. In the first-order case we regard these structures as sets of atoms built over arbitrary non-empty domains $D$; we denote by $At(D, F, P)$ the set of atomic sentences of $\langle D, F, P \rangle$ (if $D = C$, we obtain the set of atomic sentence of the language $\mathcal{L} = \langle C, F, P \rangle$);[2] and we denote by $\mathcal{T}(D, F)$ the set of ground terms of $\langle D, F, P \rangle$. If $\mathcal{L} = \langle C, F, P \rangle$ and $\mathcal{L}' = \langle C', F', P' \rangle$, we write $\mathcal{L} \subseteq \mathcal{L}'$ if $C \subseteq C'$, $F \subseteq F'$ and $P \subseteq P'$.

By an $\mathcal{L}$-interpretation over a set $D$ we mean a subset of $At(D, F, P)$. A *classical* $\mathcal{L}$-structure can be regarded as a tuple $\mathcal{I} = \langle (D, I), I^* \rangle$ where $I^*$ is an $\mathcal{L}$-interpretation over $D$ and $I \colon \mathcal{T}(C \cup D, F) \to D$, called the *assignment*, verifies that $I(d) = d$ for all $d \in D$ and is recursively defined.[3] If $D = \mathcal{T}(C, F)$ and $I = id$, $\mathcal{I}$ is known as an *Herbrand structure*. On the other hand, a *here-and-there* $\mathcal{L}$-structure with static domains, or **QHT**$^s(\mathcal{L})$-*structure*, is a tuple $\mathcal{I} = \langle (D, I), I^h, I^t \rangle$ where $\langle (D, I), I^h \rangle$ and $\langle (D, I), I^t \rangle$ are classical $\mathcal{L}$-structures such that $I^h \subseteq I^t$.

Thus we can think of a here-and-there structure $\mathcal{I}$ as similar to a first-order classical model, but having two parts, or components, $h$ and $t$ that correspond to two different points or "worlds", 'here' and 'there', in the sense of Kripke semantics for intuitionistic logic, where the worlds are ordered by $h \leq t$. At each world $w \in \{h, t\}$ one verifies a set of atoms $I^w$ in the expanded language for the domain $D$. We call the model static, since, in contrast to say intuitionistic logic, the same domain serves each of the worlds. Since $h < t$, whatever is verified at $h$ remains true at $t$. The satisfaction relation for $\mathcal{I}$ is defined so as to reflect the two different components, so we write $\mathcal{I}, w \models \varphi$ to denote that $\varphi$ is true in $\mathcal{I}$ with respect to the $w$ component. Although we only need to define the satisfaction relation in $\mathcal{L} = \langle C, P \rangle$, the recursive definition forces us to consider formulas from $\langle C \cup D, F, P \rangle$. In particular, if $p(t_1, \ldots, t_n) \in At(C \cup D, F, P)$ then $\mathcal{I}, w \models p(t_1, \ldots, t_n)$ iff $p(I(t_1), \ldots, I(t_n)) \in I^w$ for every $t_1, \ldots, t_n \in \mathcal{T}(C \cup D, F)$. Then $\models$ is extended recursively as follows[4]:

- $\mathcal{I}, w \models \varphi \wedge \psi$ iff $\mathcal{I}, w \models \varphi$ and $\mathcal{I}, w \models \psi$.

- $\mathcal{I}, w \models \varphi \vee \psi$ iff $\mathcal{I}, w \models \varphi$ or $\mathcal{I}, w \models \psi$.

- $\mathcal{I}, t \models \varphi \to \psi$ iff $\mathcal{I}, t \not\models \varphi$ or $\mathcal{I}, t \models \psi$.

- $\mathcal{I}, h \models \varphi \to \psi$ iff $\mathcal{I}, t \models \varphi \to \psi$ and $\mathcal{I}, h \not\models \varphi$ or $\mathcal{I}, h \models \psi$.

- $\mathcal{I}, w \models \neg\varphi$ iff $\mathcal{I}, t \not\models \varphi$.

- $\mathcal{I}, t \models \forall x \varphi(x)$ iff $\mathcal{I}, t \models \varphi(d)$ for all $d \in D$.

- $\mathcal{I}, h \models \forall x \varphi(x)$ iff $\mathcal{I}, t \models \forall x \varphi(x)$ and $\mathcal{I}, h \models \varphi(d)$ for all $d \in D$.

- $\mathcal{I}, w \models \exists x \varphi(x)$ iff $\mathcal{I}, w \models \varphi(d)$ for some $d \in D$.

---

2. We can think of the objects in $D$ as additional constants; this approach allow us to use a simplified notation where the objects are not distinguished from their names.

3. That is, for every $a \in C$, $I(a) \in D$ and for every $f \in F$ with arity $n$, a mapping $f^I \colon D^n \to D$ is defined; so the recursive definition is given by $I(f(t_1, \ldots, t_n)) = f^I(I(t_1), \ldots, I(t_n))$.

4. The following corresponds to the usual Kripke semantics for intuitionistic logic given our assumptions about the two worlds $h$ and $t$ and the single domain $D$,

Truth of a sentence in a model is defined as follows: $\mathcal{I} \models \varphi$ iff $\mathcal{I}, w \models \varphi$ for each $w \in \{h, t\}$. A sentence $\varphi$ is valid if it is true in all models, denoted by $\models \varphi$. A sentence $\varphi$ is a consequence of a set of sentences $\Pi$, denoted $\Pi \models \varphi$, if every model of $\Pi$ is a model of $\varphi$.

The resulting logic is called *Quantified Here-and-There Logic with static domains* (Lifschitz, Pearce, & Valverde, 2007) denoted by $\mathbf{QHT}^s$. In terms of satisfiability and validity this logic is equivalent to the logic introduced by Pearce and Valverde (2005).

A complete axiomatisation of $\mathbf{QHT}^s$ can be obtained as follows (Lifschitz et al., 2007). We take the axioms and rules of first-order intuitionistic logic and add the axiom of Hosoi

$$\alpha \vee (\neg \beta \vee (\alpha \to \beta)) \tag{5}$$

which determines 2-element here-and-there models in the propositional case, together with the axiom:

$$\exists x(\alpha(x) \to \forall x\alpha(x)).$$

We also consider the equality predicate, $\dot{=} \notin P$, interpreted by the following condition for every $w \in \{h, t\}$

- $\mathcal{I}, w \models t_1 \dot{=} t_2$    iff    $I(t_1) = I(t_2)$.

To obtain a complete axiomatisation, we then need to add the axiom of "decidable equality"

$$\forall x \forall y(x \dot{=} y \vee x \not\dot{=} y).$$

We denote the resulting logic by $\mathbf{QHT}^s_=$ (Lifschitz et al., 2007) and its inference relation by $\vdash$. By compactness a strong form of completeness can be established such that $\Pi \models \varphi$ if and only if $\Pi \vdash \varphi$.

In the context of logic programs, the following assumptions often play a role. In the case of both classical and $\mathbf{QHT}^s_=$ models, we say that the *parameter names assumption (PNA)* applies in case $I|_{\mathcal{T}(C,F)}$ is surjective, i.e. there are no unnamed individuals in D; the *unique names assumption (UNA)* applies in case $I|_{\mathcal{T}(C,F)}$ is injective; in case both the PNA and UNA apply, the *standard names assumption (SNA)* applies, i.e. $I|_{\mathcal{T}(C,F)}$ is a bijection.

As usual in first order logic, satisfiability and validity are independent of the signature. If $\mathcal{I} = \langle (D, I), I^h, I^t \rangle$ is an $\mathcal{L}'$-structure and $\mathcal{L}' \supset \mathcal{L}$, we denote by $\mathcal{I}|_\mathcal{L}$ the restriction of $\mathcal{I}$ to the sublanguage $\mathcal{L}$:

$$\mathcal{I}|_\mathcal{L} = \langle (D, I|_\mathcal{L}), I^h|_\mathcal{L}, I^t|_\mathcal{L} \rangle$$

**Proposition 2** *Suppose that $\mathcal{L}' \supset \mathcal{L}$, $\Pi$ is a theory in $\mathcal{L}$ and $\mathcal{M}$ is an $\mathcal{L}'$-model of $\Pi$. Then $\mathcal{M}|_\mathcal{L}$ is a $\mathcal{L}$-model of $\Pi$.*

**Proposition 3** *Suppose that $\mathcal{L}' \supset \mathcal{L}$ and $\varphi \in \mathcal{L}$. Then $\varphi$ is valid (resp. satisfiable) in $\mathbf{QHT}^s_=(\mathcal{L})$ if and only if is valid (resp. satisfiable) in $\mathbf{QHT}^s_=(\mathcal{L}')$.*

This proposition allows us to omit reference to the signature in the logic so it can be denoted simply by $\mathbf{QHT}^s_=$.

To simplify notation we also symbolise a $\mathbf{QHT}^s_=$ structure $\mathcal{I} = \langle (D, I), I^h, I^t \rangle$ by $\langle U, H, T \rangle$, where $U = (D, I)$ is the universe, and $H$, $T$ respectively are the sets of atoms $I^h$, $I^t$. In the case of propositional $\mathbf{HT}$ logic, Kripke structures can be regarded as pairs $\langle H, T \rangle$ of set of atoms in the obvious way. A (strongly) complete axiomatisation for $\mathbf{HT}$ is obtained from intuitionistic logic by adding just the Hosoi axiom (5).

### 2.3 Interpolation in the Logic of Here-and-There

An important and useful property of **HT** is the fact that it is the strongest propositional intermediate logic (i.e. strengthening of intuitionistic logic) that is properly contained in classical logic. Moreover it in turn properly contains all other such intermediate logics. In addition **HT** is one of precisely seven superintuitionistic propositional logics possessing the interpolation property (Maksimova, 1977; Gabbay & Maksimova, 2005).

For languages without function symbols Ono showed that interpolation holds in the logic $\mathbf{QHT}^s$ of quantified here-and-there with constant domains (Ono, 1983).[5] In addition, Maksimova (1997, 1998) showed that adding pure equality axioms, e.g. the decidable equality axiom, to any superintuitionistic logic preserves the interpolation property (Gabbay & Maksimova, 2005). We conclude therefore

**Proposition 4** *The logic $\mathbf{QHT}^s_{\doteq}$ possesses the interpolation property.*

Note that by the strong completeness theorem for $\mathbf{QHT}^s_{\doteq}$ we can work equivalently with $\vdash$ or with $\models$.

Here we can make the further observation that interpolation continues to hold for languages that include function symbols. This can be established using the following property.

**Proposition 5** *For every formula $\varphi$, it is possible to build a formula $\psi$, such that $\varphi \equiv \psi$, and the atoms of $\psi$ are of one of the following types:*

- $x \doteq a$ *for some $a \in C$,*

- $f(x_1, \ldots x_n) \doteq y$ *for some $f \in F$ (where every $x_i$ and $y$ are variables),*

- $p(t_1, \ldots, t_m)$ *(where every $x_i$ and $y$ are variables).*

**Theorem 1** *Let $\mathcal{L}$ be a language containing function symbols. Then $\mathbf{QHT}^s_{\doteq}(\mathcal{L})$ has the interpolation property.*

*Proof sketch:* Let us assume that $\vdash \varphi \to \psi$; from the previous proposition, we can assume, without loss of generality, that the function symbols in $\varphi$ and $\psi$ are in atoms of type $f(x_1, \ldots x_n) \doteq y$. Now, we consider a language $\mathcal{L}'$ obtained from $\mathcal{L}$ by replacing every function symbol $f$ by a fresh predicate symbol, $P_f$, such that the $\mathrm{Arity}(P_f) = 1 + \mathrm{Arity}(f)$. Let $\varphi'$ and $\psi'$ be formulas in $\mathcal{L}'$ build from $\varphi$ and $\psi$ respectively, by replacing every atom $f(x_1, \ldots x_n) \doteq y$ by $P_f(x_1, \ldots x_n, y)$. Trivially, $\vdash \varphi' \to \psi'$ and, for the interpolation property of $\mathbf{QHT}^s_{\doteq}(\mathcal{L}')$, there exists an interpolant $\beta'$: $\vdash \varphi' \to \beta'$, $\vdash \beta' \to \psi'$. If we replace in $\beta'$ the predicates $P_f(t_1, \ldots, t_n, t_{n+1})$ by atoms $f(t_1, \ldots t_n) \doteq t_{n+1}$ we obtain the interpolant $\beta$ for the initial pair of formulas. $\square$

---

5. Ono's axiomatisation of $\mathbf{QHT}^s$ uses the constant domains axiom $\forall x(\alpha(x) \lor \beta) \to (\forall x \alpha(x) \lor \beta)$, as well as alternative axioms for propositional here-and-there, viz. $p \lor (p \to (q \lor \neg q))$ and $(p \to q) \lor (q \to p) \lor (p \leftrightarrow \neg q)$. However, the axioms given here are equivalent to Ono's.

## 2.4 Equilibrium Logic

Equilibrium logic is a non-monotonic logic based on certain kinds of minimal models in $\mathbf{QHT}^s_=$ or $\mathbf{HT}$. We give the definition for $\mathbf{QHT}^s_=$; the propositional version is easily obtained from it.

**Definition 5** *Among quantified here-and-there structures we define the order $\trianglelefteq$ as follows:*

$$\langle (D,I), I^h, I^t \rangle \trianglelefteq \langle (D',J), J^h, J^t \rangle \qquad if \qquad D=D',\ I=J,\ I^t=J^t,\ I^h \subseteq J^h.$$

*If the subset relation holds strictly, we write '$\triangleleft$'.*

**Definition 6 (Equilibrium model)** *Let $\Pi$ be a theory and $\mathcal{I} = \langle (D,I), I^h, I^t \rangle$ a model of $\Pi$.*

1. *$\mathcal{I}$ is said to be* total *if $I^h = I^t$.*

2. *$\mathcal{I}$ is said to be an* equilibrium *model of $\Pi$ (or short, we say: "$\mathcal{I}$ is in equilibrium") if it is minimal under $\trianglelefteq$ among models of $\Pi$, and it is total.*

In other words, equilibrium models are total models for which there is no 'smaller' non-total model. Evidently a total $\mathbf{QHT}^s_=$ model of a theory $\Gamma$ can be equivalently regarded as a classical first order model of $\Gamma$; and in what follows we make tacit use of this equivalence. In the propositional case, equilibrium models are defined in the same way, where now the ordering is between propositional $\mathbf{HT}$ models. In the usual way a formula or theory is said to be *consistent* if it has a $\mathbf{QHT}^s_=$ model and additionally we say that it is *coherent* if it has an equilibrium model.

The following definition give a preliminary notion of equilibrium entailment, which agrees with standard versions of equilibrium logic (Pearce, 2006).

**Definition 7** *The relation $\mathrel|\!\sim$, called* equilibrium entailment*, is defined as follows. Let $\Pi$ be a set of formulas.*

1. *If $\Pi$ is non-empty and coherent (has equilibrium models), then $\Pi \mathrel|\!\sim \varphi$ if every equilibrium model of $\Pi$ is a model of $\varphi$ in $\mathbf{QHT}^s_=$ (respectively $\mathbf{HT}$).*

2. *If either $\Pi$ is empty or has no equilibrium models, then $\Pi \mathrel|\!\sim \varphi$ if $\Pi \vdash \varphi$.*

Notice that unless we need to distinguish propositional from first-order reasoning we use the symbols '$\mathrel|\!\sim$', '$\vdash$' and '$\models$' for either version.

A few words may help to explain the concept of equilibrium entailment. First, we define the basic notion of entailment as truth in every intended (equilibrium) model. In nonmonotonic reasoning this is a common approach and sometimes called a *skeptical* or *cautious* notion of entailment or inference; its counterpart *brave* reasoning being defined via truth in some intended model. Since equilibrium logic is intended to provide a logical foundation for the answer set semantics of logic programs, the cautious variant of entailment is the natural one to choose: the standard consequence relation associated with answer sets is given by truth in all answer sets of a program. Note however that in ASP as a programming paradigm each answer set may correspond to a particular solution of the problem being modelled and is therefore of interest in its own right.

Secondly, it is useful to have a nonmonotonic consequence or entailment relation that is non-trivially defined for all consistent theories. As we shall see below, however, not all such theories possess equilibrium models. For such cases it is natural to use monotonic consequence as the entailment relation. In particular in the propositional case **HT** is a maximal logic with the property that logically equivalent theories have the same equilibrium models. Evidently situation 2 also handles correctly the cases that $\Pi$ is empty or inconsistent.

Despite these qualifications, there remains an ambiguity in the concept of equilibrium entailment that we now need to settle. Suppose that $\mathcal{L}' \supset \mathcal{L}$, $\Pi$ is a theory in $\mathcal{L}$ and $\varphi$ is a sentence in $\mathcal{L}'$ (i.e. $\mathcal{L}' = \mathcal{L}(\varphi)$). How should we understand the expression '$\Pi \hspace{0.1em}\vdash\hspace{-0.55em}\sim \varphi$'?

Evidently, if we fix a language in advance, say as the language $\mathcal{L}'$, then we can simply consider the equilibrium models of $\Pi$ in $\mathcal{L}'$. But if $\Pi$ represents a knowledge base or a logic program, for instance, we may also take the view that $\mathcal{L}(\Pi)$ is the appropriate signature to work with. In that case, the query $\varphi$ is as such not fully interpreted as it contains some terms not in the theory language $\mathcal{L}(\Pi)$.

For any language $\mathcal{L}$ and theory $\Pi$ whose language is contained in $\mathcal{L}$, let $EM_{\mathcal{L}}(\Pi)$ be the collection of all equilibrium models of $\Pi$ in $\mathbf{QHT}^s_=(\mathcal{L})$. Now consider the following two variants of entailment.

**Definition 8 (Equilibrium entailment)** *Assume $\Pi$ is a theory in $\mathcal{L}$, is non-empty and has equilibrium models, then:*
*(i) Let us write $\Pi \hspace{0.1em}\vdash\hspace{-0.55em}\sim_{cw} \varphi$ if and only if $\mathcal{M} \models \varphi$ for each $\mathcal{M} \in EM_{\mathcal{L}'}(\Pi)$, where $\mathcal{L}' = \mathcal{L} \cup \mathcal{L}(\varphi)$:*
*(ii) let us write $\Pi \hspace{0.1em}\vdash\hspace{-0.55em}\sim_{ow} \varphi$ if and only if $\mathcal{M} \models \varphi$ for each $\mathcal{M} \in EM_{\mathcal{L}}(\Pi) \upharpoonright^{\mathcal{L}(\varphi)}$, where in general $EM_{\mathcal{L}}(\Pi) \upharpoonright^{\mathcal{L}'}$ denotes the collection of all expansions of elements of $EM_{\mathcal{L}}(\Pi)$ to models in $\mathcal{L} \cup \mathcal{L}'$, i.e. where the vocabulary of $\mathcal{L}' \setminus \mathcal{L}$ is interpreted arbitrarily.*
*(iii) If either $\Pi$ is empty or has no equilibrium models, then $\Pi \hspace{0.1em}\vdash\hspace{-0.55em}\sim_{cw} \varphi$ iff $\Pi \hspace{0.1em}\vdash\hspace{-0.55em}\sim_{ow} \varphi$ iff $\Pi \vdash \varphi$.*

A simple example will illustrate the difference between $\hspace{0.1em}\vdash\hspace{-0.55em}\sim_{cw}$ and $\hspace{0.1em}\vdash\hspace{-0.55em}\sim_{ow}$. Let $\psi$ be an $\mathcal{L}$-sentence and let $q(x)$ be a predicate not in $\mathcal{L}$. Let $a$ be a constant in $\mathcal{L}$ and let $\mathcal{L}'$ be the language $\mathcal{L} \cup \{q\}$. By the first method we have $\psi \hspace{0.1em}\vdash\hspace{-0.55em}\sim_{cw} \psi \wedge (q(a) \vee \neg q(a))$. In fact we have the stronger entailment $\psi \hspace{0.1em}\vdash\hspace{-0.55em}\sim_{cw} \psi \wedge \neg q(a)$. The reason is that when we form the equilibrium models of $\psi$ in $\mathcal{L}'$, $q(a)$ will be false in each as an effect of taking minimal models. On the other hand, if we expand equilibrium models of $\psi$ in $\mathbf{QHT}^s_=(\mathcal{L})$ to $\mathbf{QHT}^s_=(\mathcal{L}')$, the new predicate $q$ receives an arbitrary interpretation in $\mathbf{QHT}^s_=(\mathcal{L}')$. Since this logic is 3-valued we do not obtain $\Pi \hspace{0.1em}\vdash\hspace{-0.55em}\sim_{ow} q(a) \vee \neg q(a)$.

For standard, monotonic logics, there is no difference between these two forms of entailment. If in Definition 8 we replace everywhere equilibrium model by simply model (in $\mathbf{QHT}^s_=$), variants (i) and (ii) give the same result.

In the context of logic programming and deductive databases the more orthodox view is that reasoning is based on a *closed world assumption* (CWA). Accordingly a ground atomic query like $q(a)$?, where the predicate $q$ does not belong to the language of the program or database, would simply be assigned the value *false*. This is also the case with the first kind of equilibrium entailment and we use the label $\hspace{0.1em}\vdash\hspace{-0.55em}\sim_{cw}$ since this variant appears closer to a closed world form of reasoning. On the other hand, there may be legitimate cases where we do not want to apply the CWA and where unknown values should be assigned to an atom that is not expressed in the theory language. Then the second form of entailment, $\hspace{0.1em}\vdash\hspace{-0.55em}\sim_{ow}$,

nearer to open world reasoning, may be more appropriate. For present purposes, however, the suffices '*cw*' and '*ow*' should be thought of merely as mnemonic labels.

A thorough analysis of closed world versus open world reasoning in this context would lead us to consider assumptions such as UNA or SNA and is outside the scope of this paper. However, it has been observed in logic programming that the use of CWA can lead to certain apparent anomalies. Notably this occurs with programs that are *unsafe* (see Section 5 below), such as the following, formulated in traditional notation for logic programs:[6]

$q(x, y) :-\ not\ p(x, y).$

$p(x, x).$

Given restrictions such as SNA or Herbrand models, the query

$? - q(a, z).$

yields no answer for $z$ because it cannot be satisfied in models with only a single domain element $a$, while the query

$? - q(a, b).$

is satisfiable, given the new constant $b$. In logic programming, where these restrictions are usually assumed, different solutions to this problem have been proposed (Gelder, Ross, & Schlipf, 1991; Kunen, 1987; Maher, 1988). Here we would like to point out that for equilibrium logic generally speaking this kind of program or theory does not create any special difficulties. Neither the query

$? - q(a, z).$

which is understood as $\exists z q(a, z)$, nor the query

$? - q(a, b).$

is true in all equilibrium models. In particular, in an equilibrium model whose domain is a singleton element, even $q(a, b)$ need not be true; evidently in the general case that UNA for instance does not apply. On the other hand in answer set programming, where UNA is often assumed, it is also typically assumed that programs are *safe*. By the safety condition the above program is excluded because variables appearing in the head of a rule do not appear in its positive body and this makes answer sets sensitive to the set of constants appearing in the language or those that are used for grounding the program. In this paper, where the application of interpolation in ASP is concerned, we restrict attention to safe programs and theories complying with a generalised form of safety (Section 5 below).

## 3. Interpolation in Propositional Equilibrium Logic

In this section we deal with interpolation in propositional equilibrium logic. It is clear that by its semantic construction propositional equilibrium logic has **HT** as a deductive base. This base is actually maximal.

**Proposition 6 HT** *is a strong and maximal deductive base for (propositional) equilibrium entailment.*

The first property is precisely the strong equivalence theorem of Lifschitz, Pearce and Valverde (2001). Maximality follows from the fact that any logic strictly stronger than **HT** would have to contain classical logic which is easily seen not to be a deductive base, e.g. violating condition (ii) of Definition 1. We have:

---

6. We are grateful to an anonymous referee for raising this point and the example.

**Lemma 1** *Let $\alpha$ be a coherent $\mathbf{HT}$-formula and $EM(\alpha)$ its set of equilibrium models. Then there is formula $\alpha'$ of $\mathbf{HT}$ in $v(\alpha)$ that defines $EM(\alpha)$ in the sense that $\mathcal{M} \in EM(\alpha)$ if and only if $\mathcal{M} \models \alpha'$.*

*Proof.* Suppose that $\alpha$ is coherent. and let

$$\mathcal{M}_1 = \langle T_1, T_1 \rangle, \quad \mathcal{M}_2 = \langle T_2, T_2 \rangle, \dots, \ \mathcal{M}_n = \langle T_n, T_n \rangle$$

be an enumeration of its equilibrium models. We show how to define $EM(\alpha)$. Suppose each $T_i$, has $k_i$ elements and denote them by $A_1^i, \dots, A_j^i, \dots, A_{k_i}^i$. Let $\overline{T_i}$ be the complement of $T_i$; then we can list its members as $A_{k_i+1}^i, \dots A_l^i \dots, A_{|v(\alpha)|}^i$. Set

$$\delta^i = \bigwedge_{j=1,\dots,k_i} A_j^i \wedge \neg( \bigvee_{l=k_{i+1},\dots,|v(\alpha)|} A_l^i), \qquad \text{and} \quad \alpha' = \bigvee_{i=1,\dots,n} \delta^i$$

We claim that $\mathcal{M} \models \alpha'$ if and only if $\mathcal{M} = \mathcal{M}_i$ for some $i = 1, \dots, n$, i.e. the models of $\alpha'$ are precisely $\mathcal{M}_1, \dots, \mathcal{M}_n$. To verify this claim, note that each $\mathcal{M}_i \models \delta^i$ and so $\mathcal{M}_i \models \alpha'$. Conversely, suppose that $\mathcal{M} \models \alpha'$. From the semantics of $\mathbf{HT}$ it is clear that $\mathcal{M} \models \varphi \vee \psi$ iff $\mathcal{M} \models \varphi$ or $\mathcal{M} \models \psi$, so in particular $\mathcal{M} \models \alpha'$ implies $\mathcal{M} \models \delta^i$ for some $i = 1, \dots, n$. However, each $\delta^i$ defines a complete theory whose models are total. It follows that if $\mathcal{M} \models \delta^i$, then $\mathcal{M} = \mathcal{M}_i$. This establishes the claim. □

Although we shall now demonstrate interpolation in the $(\hspace{1pt}\vert\!\sim, \hspace{1pt}\vert\!\sim)$ form for the relation $\hspace{1pt}\vert\!\sim_{cw}$, we actually establish a stronger result. One consequence of this is that if we are concerned with $\hspace{1pt}\vert\!\sim_{ow}$ entailment then the $(\hspace{1pt}\vert\!\sim, \vdash)$ form of interpolation actually holds.

**Proposition 7** ($\hspace{1pt}\vert\!\sim, \hspace{1pt}\vert\!\sim$**-Interpolation**) *Let $\alpha, \beta$ be formulas and set $v = v(\alpha) \cup v(\beta)$ and $v' = v(\beta) \setminus v(\alpha)$ and suppose that $B_1, \dots B_n$ is an enumeration of $v'$. If $\alpha \hspace{1pt}\vert\!\sim_{cw} \beta$, there is a formula $\gamma$ such that $v(\gamma) \subseteq v(\alpha) \cap v(\beta)$, $\alpha \hspace{1pt}\vert\!\sim \gamma$, and $\gamma \wedge \neg B_1 \wedge \dots \wedge \neg B_n \models \beta$. Hence in particular $\gamma \hspace{1pt}\vert\!\sim_{cw} \beta$.*

*Proof.* Let $\alpha, \beta$ and $v, v'$ be as in the statement of the proposition, and suppose that $\alpha \hspace{1pt}\vert\!\sim_{cw} \beta$. Then $\beta$ holds in all equilibrium models of $\alpha$ in the language $v$. Case (i): suppose that $\alpha$ is coherent and form its set of equilibrium models, $EM_v(\alpha)$.

By the equilibrium construction it is easy to see that in each model $\mathcal{M} \in EM_v(\alpha)$ each atom $B_i$ is false, for $i = 1, n$. Construct the formulas $\delta_i$ and the formula $\alpha'$ exactly as in the proof of Lemma 1. Now consider the formula $(\neg B_1 \wedge \dots \wedge \neg B_n) \wedge \alpha'$. Clearly this formula defines the set of equilibrium models of $\alpha$ in $\mathbf{HT}(v)$. Consequently, $(\neg B_1 \wedge \dots \wedge \neg B_n) \wedge \alpha' \models \beta$ and so $\alpha' \vdash (\neg B_1 \wedge \dots \wedge \neg B_n) \to \beta$. We can now apply the interpolation theorem for $\mathbf{HT}$ to infer that there is a formula $\gamma$ such that $\alpha' \vdash \gamma$ and $\gamma \vdash (\neg B_1 \wedge \dots \wedge \neg B_n) \to \beta$, where $v(\gamma) \subseteq v(\alpha') \cap v(\beta)$ and hence $v(\gamma) \subseteq v(\alpha) \cap v(\beta)$. Since $\mathbf{HT}$ is a deductive base, we conclude that

$$\alpha \hspace{1pt}\vert\!\sim \gamma \ \ \& \ \ \gamma \wedge \neg B_1 \wedge \dots \wedge \neg B_n \vdash \beta.$$

Now, since $v(\gamma) \subseteq v(\alpha) \cap v(\beta)$, $B_i \notin v(\gamma)$ for $i = 1, n$. It follows that in $\mathbf{HT}(v(\beta))$, each $B_i$ is false in every equilibrium model of $\gamma$. So each such model $\mathcal{M}$ satisfies $(\neg B_1 \wedge \dots \wedge \neg B_n)$.[7] Since each also satisfies $\beta$, we have $\gamma \hspace{1pt}\vert\!\sim_{cw} \beta$.

---

7. Notice that in this case adding to $\gamma$ the sentence $(\neg B_1 \wedge \dots \wedge \neg B_n)$ does not change its set of equilibrium models.

Case (ii). If $\alpha$ has no equilibrium models then the hypothesis is that $\alpha \vdash \beta$. In that case we simply choose an interpolant $\gamma$ for $(\alpha, \beta)$. $\qquad\square$

**Corollary 1** ($\hspace{0.1em}\vdash\hspace{-0.75em}\sim, \vdash$-**Interpolation**) *Let $\alpha, \beta$ be formulas such that $\alpha \hspace{0.1em}\vdash\hspace{-0.75em}\sim_{cw} \beta$ and $v(\beta) \subseteq v(\alpha)$. There is a formula $\gamma$ such that $v(\gamma) \subseteq v(\alpha) \cap v(\beta)$ and $\alpha \hspace{0.1em}\vdash\hspace{-0.75em}\sim_{cw} \gamma$ and $\gamma \vdash \beta$.*

*Proof.* Immediate from Proposition 7 by the fact that $v(\beta) \setminus v(\alpha) = \varnothing$. $\qquad\square$

**Proposition 8** ($\hspace{0.1em}\vdash\hspace{-0.75em}\sim, \vdash$-**Interpolation**) *Let $\alpha, \beta$ be formulas and set $v = v(\alpha) \cup v(\beta)$ and $v' = v(\beta) \setminus v(\alpha)$. If $\alpha \hspace{0.1em}\vdash\hspace{-0.75em}\sim_{ow} \beta$, there is a formula $\gamma$ such that $v(\gamma) \subseteq v(\alpha) \cap v(\beta)$, $\alpha \hspace{0.1em}\vdash\hspace{-0.75em}\sim \gamma$, and $\gamma \vdash \beta$.*

*Proof.* Let $\alpha, \beta$ and $v, v'$ be as in the statement of the proposition and suppose that $\alpha \hspace{0.1em}\vdash\hspace{-0.75em}\sim_{ow} \beta$. Then $\beta$ holds in all expansions of elements of $EM_{v(\alpha)}(\alpha)$ to the language $v$. Case (i): suppose that $\alpha$ is coherent and consider $EM_{v(\alpha)}(\alpha)$.

Again construct the formulas $\delta_i$ and the formula $\alpha'$ exactly as in the proof of Lemma 1. Now consider $\alpha'$ which defines the set $EM_{v(\alpha)}(\alpha)$. Then $\beta$ holds in all expansions of models of $\alpha'$ to $v$. Hence $\alpha' \models \beta$ and therefore $\alpha' \vdash \beta$ We can now apply the interpolation theorem for **HT** to infer that there is a formula $\gamma$ such that $\alpha' \vdash \gamma$ and $\gamma \vdash \beta$, where $v(\gamma) \subseteq v(\alpha') \cap v(\beta)$ and hence $v(\gamma) \subseteq v(\alpha) \cap v(\beta)$. Since $\alpha \hspace{0.1em}\vdash\hspace{-0.75em}\sim_{ow} \alpha'$ and **HT** is a deductive base we conclude that

$$\alpha \hspace{0.1em}\vdash\hspace{-0.75em}\sim_{ow} \gamma \;\;\&\;\; \gamma \vdash \beta.$$

Case (ii). If $\alpha$ has no equilibrium models, choose $\gamma$ as an interpolant for $(\alpha, \beta)$. $\qquad\square$

## 4. Interpolation in Quantified Equilibrium Logic

We now turn to first-order logic. Given inferences of the form $\alpha \hspace{0.1em}\vdash\hspace{-0.75em}\sim \beta$, a key element in the proofs of Propositions 7 and 8 is the existence of a formula $\alpha'$ that defines the collection $EM_{v(\alpha)}(\alpha)$ of equilibrium models. In the propositional case we have seen how the existence of such an $\alpha'$ can be established. In the first-order case, on the other hand, such an $\alpha'$ need not exist. In other words, $EM_{\mathcal{L}(\alpha)}(\alpha)$ need not be first-order definable for arbitrary $\alpha$. This fact is not hard to show. As Ferraris et al. (2007) have pointed out, in the general form of answer set programming where first-order formulas are allowed, and *a fortiori* in quantified equilibrium logic, the property of *transitive closure* is expressible. Yet this property is not definable in classical first-order logic and therefore it also cannot be defined in $\mathbf{QHT}^s_=$.

In the usual way we say that a collection $\mathcal{K}$ of $\mathbf{QHT}^s_=(\mathcal{L})$ models is ($\mathbf{QHT}^s_=$) definable if there is an $\mathcal{L}$-sentence, $\varphi$, such that $\mathcal{M} \in \mathcal{K} \Leftrightarrow \mathcal{M} \models \varphi$. It is easy to see that whenever the class $EM_{\mathcal{L}(\alpha)}(\alpha)$ is first-order definable in $\mathbf{QHT}^s_=$ we do obtain first-order analogs of Propositions 7 and 8. The method of proof is essentially the same as before. For completeness we outline the main steps for the case of ($\hspace{0.1em}\vdash\hspace{-0.75em}\sim, \hspace{0.1em}\vdash\hspace{-0.75em}\sim$)-interpolation.

**Proposition 9** ($\hspace{0.1em}\vdash\hspace{-0.75em}\sim, \hspace{0.1em}\vdash\hspace{-0.75em}\sim$-**Interpolation**) *Let $\alpha, \beta$ be formulas such that the collection of equilibrium models of $\alpha$ is $\mathbf{QHT}^s_=$- definable. Set $\mathcal{L} = \mathcal{L}(\alpha) \cup \mathcal{L}(\beta)$ and $\mathcal{L}' = \mathcal{L}(\beta) \setminus \mathcal{L}(\alpha)$. Let $\{p_i : i = 1, n\}$ be the (finite, possibly empty) set of predicates in $\mathcal{L}'$ and suppose for each $i$* that $p_i$ is of arity $k_i$. If $\alpha \mathrel{|\!\sim}_{cw} \beta$, there is a formula $\gamma$ such that $\mathcal{L}(\gamma) \subseteq \mathcal{L}(\alpha) \cap \mathcal{L}(\beta)$, $\alpha \mathrel{|\!\sim} \gamma$, and

$$\gamma \wedge \bigwedge_{i=1,n} \forall \mathbf{x} \neg p_i(\mathbf{x}) \models \beta$$

Hence in particular $\gamma \mathrel{|\!\sim}_{cw} \beta$.

*Proof.* Assume the hypotheses. Then $\beta$ holds in all equilibrium models of $\alpha$ in the language $\mathcal{L}$. We treat just the case where $\alpha$ is coherent and has a non-empty collection of equilibrium models, $EM_{\mathcal{L}(\alpha)}(\alpha)$. By assumption this collection is definable by a $\mathbf{QHT}^s_=(\mathcal{L}(\alpha))$-sentence, $\alpha'$, say. Now consider the equilibrium models of $\alpha$ in the expanded language $\mathcal{L}$, i.e. the collection $EM_{\mathcal{L}}(\alpha)$. By the equilibrium construction we claim that $EM_{\mathcal{L}}(\alpha) \models \forall \mathbf{x} \neg p_i(\mathbf{x})$, for all $i = 1, n$. Since we are now working with the first-order semantics, let us rehearse briefly the argument for this. If it were not true there would be a model $\langle U, T, T \rangle \in EM_{\mathcal{L}}(\alpha)$, a predicate symbol $p_i \in \mathcal{L}'$ and some tuple $\mathbf{a}$ of elements in the domain of $\langle U, T, T \rangle$, such that $\langle U, T, T \rangle \models p_i(\mathbf{a})$, ie $p_i(\mathbf{a}) \in T$. However, since $\alpha$ does not refer to the relation $p_i$, the structure $\langle U, H, T \rangle$ with $H = T \setminus p_i(\mathbf{a})$ must also be a model of $\alpha$, contradicting that $\langle U, T, T \rangle$ is in equilibrium. So $EM_{\mathcal{L}}(\alpha) \models \alpha' \wedge \bigwedge_{i=1,n} \forall \mathbf{x} \neg p_i(\mathbf{x})$ and since $\alpha'$ defines $EM_{\mathcal{L}(\alpha)}(\alpha)$ clearly $\alpha' \wedge \bigwedge_{i=1,n} \forall \mathbf{x} \neg p_i(\mathbf{x})$ defines $EM_{\mathcal{L}}(\alpha)$.

Now we proceed as in the propositional case. $\alpha' \wedge \bigwedge_{i=1,n} \forall \mathbf{x} \neg p_i(\mathbf{x}) \vdash \beta$, so

$$\alpha' \vdash \bigwedge_{i=1,n} \forall \mathbf{x} \neg p_i(\mathbf{x}) \to \beta.$$

By the interpolation theorem for $\mathbf{QHT}^s_=$ there is a formula $\gamma$ such that $\mathcal{L}(\gamma) \subseteq \mathcal{L}(\alpha) \cap \mathcal{L}(\beta)$, $\alpha' \vdash \gamma$ and $\gamma \vdash \bigwedge_{i=1,n} \forall \mathbf{x} \neg p_i(\mathbf{x}) \to \beta$. Consequently we also have

$$\alpha \mathrel{|\!\sim} \gamma \ \ \& \ \ \gamma \wedge \bigwedge_{i=1,n} \forall \mathbf{x} \neg p_i(\mathbf{x}) \vdash \beta$$

By the same token as in the propositional case, we infer that $\gamma \mathrel{|\!\sim}_{cw} \beta$. $\qquad\square$

The case of $(\mathrel{|\!\sim}, \vdash)$-interpolation for $\mathrel{|\!\sim}_{ow}$ is analogous and we state the main property without proof.

**Proposition 10 ($\mathrel{|\!\sim}, \vdash$-Interpolation)** *Let $\alpha, \beta$ be formulas such that the collection of equilibrium models of $\alpha$ is $\mathbf{QHT}^s_=$- definable. If $\alpha \mathrel{|\!\sim}_{ow} \beta$, there is a formula $\gamma$ such that $\mathcal{L}(\gamma) \subseteq \mathcal{L}(\alpha) \cap \mathcal{L}(\beta)$, $\alpha \mathrel{|\!\sim} \gamma$ and $\gamma \vdash \beta$.*

## 5. An Illustration: Interpolation for Safe Formulas

How restrictive is the definability assumption? Is it often met in practice? Actually in mainstream answer set programming, whose language equilibrium logic captures and extends (see the next section), non-definable classes of answers sets play no significant role. The reason is that for query answering existing solvers rely on grounders that instantiate all or parts of a program before computing the intended models or solutions. The grounding process essentially eliminates variables and reduces the original program to propositional form. In such practical cases, then, the collection of stable or equilibrium models will be definable.

For this computational approach to work in general, syntactic restrictions need to be imposed on admissible programs or theories. The most common form of restriction is called *safety*. For standard types of logic programs based on *rules* one regards a rule as safe if every variable appearing in rule's head also appears in its body. For the more complex formulas admitted by equilibrium logic and by the general approach to answer sets (Ferraris et al., 2007; Ferraris, 2008), new concepts of safety need to be devised. Proposals for suitable safety concepts were made by Lee, Lifschitz and Palla (2008b) for general first-order formulas and by Bria, Faber and Leone (2008) for a more restricted syntactic class. More recently Cabalar, Pearce and Valverde (2009) have generalised both these approaches and suggested a safety concept for arbitrary function-free formulas in equilibrium logic. Since this new concept of safety defines a quite broad class of interpolable formulas, let us review here its main features. In the following section we will mention some other kinds of interpolable formulas that may arise in answer set programming.

## 5.1 General Concept of Safety

For the remainder of this section we assume that all languages are function-free. As usual a sentence is said to be in *prenex* form if it has the following shape, for some $n \geq 0$:

$$Q_1 x_1 \ldots Q_n x_n \alpha$$

where $Q_i$ is $\forall$ or $\exists$ and $\alpha$ is quantifier-free. A sentence is said to be *universal* if it is in prenex form and all quantifiers are universal. A universal theory is a set of universal sentences. The safety concept is defined for prenex formulas which provide a normal form for $\mathbf{QHT}^s_=$ (Pearce & Valverde, 2005).

We first introduce a concept called semi-safety. The main property of semi-safety formulas will be that their equilibrium models only refer to objects from their language. Note that for the remainder of this section we use the fact that negation can be treated as a defined operator, by $\neg \varphi \equiv \varphi \to \bot$, and do not consider additional semantic clauses for negation.

**Definition 9 (Semi-safety)** *A quantifier free formula $\varphi$ is semi-safe it is has not non-semi-safe variable; that is, $\mathrm{NSS}(\varphi) = \varnothing$, where the NSS operator is recursively defined as follows:*

- *If $\varphi$ is an atom, $\mathrm{NSS}(\varphi)$ is the set of variables in $\varphi$;*

- *$\mathrm{NSS}(\varphi_1 \land \varphi_2) = \mathrm{NSS}(\varphi_1) \cup \mathrm{NSS}(\varphi_2)$;*

- *$\mathrm{NSS}(\varphi_1 \lor \varphi_2) = \mathrm{NSS}(\varphi_1) \cup \mathrm{NSS}(\varphi_2)$;*

- *$\mathrm{NSS}(\varphi_1 \to \varphi_2) = \mathrm{NSS}(\varphi_2) \smallsetminus \mathrm{RV}(\varphi_1)$,*

*where operator RV computes the* restricted variables *as follows:*

- *For atomic $\varphi$, if $\varphi$ is an equality between two variables then $\mathrm{RV}(\varphi) = \varnothing$; otherwise, $\mathrm{RV}(\varphi)$ is the set of all variables occurring in $\varphi$;*

- *$\mathrm{RV}(\bot) = \varnothing$;*

- $\mathrm{RV}(\varphi_1 \wedge \varphi_2) = \mathrm{RV}(\varphi_1) \cup \mathrm{RV}(\varphi_2)$;

- $\mathrm{RV}(\varphi_1 \vee \varphi_2) = \mathrm{RV}(\varphi_1) \cap \mathrm{RV}(\varphi_2)$;

- $\mathrm{RV}(\varphi_1 \to \varphi_2) = \varnothing$.

This definition of semi-safe formulas was introduced by Cabalar, Pearce and Valverde (2009) and generalises the former definition of Lee, Lifschitz and Palla (2008b). In short, a variable $x$ is semi-safe in $\varphi$ if every occurrence is inside some subformula $\alpha \to \beta$ such that, either $x \in \mathrm{RV}(\alpha)$ or $x$ is semi-safe in $\beta$.

Some examples of semi-safe formulas are, for instance:

$$\neg p(x) \to (q(x) \to r(x))$$
$$p(x) \vee q \to \neg r(x) \tag{6}$$

Note how in (6), $x$ is not restricted in $p(x) \vee q$ but the consequent $\neg r(x)$ is semi-safe and thus the formula itself. On the contrary, the following formulas are not semi-safe:

$$p(x) \vee q \to r(x)$$
$$\neg\neg p(x) \wedge \neg r(x) \to q(x)$$

The following results set the previously referred property for semi-safe formulas: their equilibrium models only include objects from the language.

**Proposition 11 (Cabalar et al., 2009)**
*If $\varphi$ is function free, semi-safe, and $\langle (D, I), T, T \rangle \models \varphi$, then $\langle (D, I), T|_C, T \rangle \models \varphi$.*

**Theorem 2 (Cabalar et al., 2009)** *If $\varphi$ is function free, semi-safe, and $\langle (D, I), T, T \rangle$ is an equilibrium model of $\varphi$, then $T|_C = T$.*

The equilibrium models of semi-safe formulas only refer to objects from the language, however a model could be or not in equilibrium depending of the considered domain. To guarantee the independence from the domain, we need an additional property to the semi-safety. Specifically, we need to analyse whether the unnamed elements could modify an interpretation of the formula. To do that, we use the assignments of the Kleene's three-valued logic; the three-valued interpretation $\nu \colon At \to \{0, 1/2, 1\}$, are extended to evaluate arbitrary formulas $\nu(\varphi)$ as follows:

$$
\begin{array}{rclcrcl}
\nu(\varphi \wedge \psi) & = & \min(\nu(\varphi), \nu(\psi)) & & \nu(\bot) & = & 0 \\
\nu(\varphi \vee \psi) & = & \max(\nu(\varphi), \nu(\psi)) & & \nu(\varphi \to \psi) & = & \max(1 - \nu(\varphi), \nu(\psi))
\end{array}
$$

For every variable $x$, we are going to use the Kleene's interpretations $\nu_x$, defined as follows: $\nu_x(\alpha) = 0$ if $x$ occurs in the atom $\alpha$ and $\nu_x(\alpha) = 1/2$ otherwise. Intuitively, $\nu_x(\varphi)$ fixes all atoms containing the variable $x$ to 0 (falsity) leaving all the rest undefined and then evaluates $\varphi$ using Kleene's three-valued operators, that is nothing else but exploiting the defined values 1 (true) and 0 (false) as much as possible.

An occurrence of a variable $x$ in $Qx\,\varphi$ is *weakly-restricted* if it occurs in a subformula $\psi$ of $\varphi$ such that:

- $Q = \forall$, $\psi$ is positive[8] and $\nu_x(\psi) = 1$

- $Q = \forall$, $\psi$ is negative and $\nu_x(\psi) = 0$

- $Q = \exists$, $\psi$ is positive and $\nu_x(\psi) = 0$

- $Q = \exists$, $\psi$ is negative and $\nu_x(\psi) = 1$

In all cases, we further say that *$\psi$ makes the ocurrence weakly restricted in $\varphi$*. This property is added to the semi-safety condition to complete the definition of safety.

**Definition 10** *A semi-safe sentence is said to be* safe *if all its positive occurrences of universally quantified variables, and all its negative occurrences of existentially quantified variables are weakly restricted.*

For instance, the formula $\varphi = \forall x(\neg q(x) \to (r \vee \neg p(x)))$ is safe: the occurrence of $x$ in $p(x)$ is negative, whereas the occurrence in $q(x)$ is inside a positive subformula, $\varphi$ itself, for which $x$ is weakly-restricted, since $\nu_x(\varphi) = \neg 0 \to (\frac{1}{2} \vee \neg 0) = 1$. Another example of a safe formula is $\forall x((\neg\neg p(x) \wedge q(x)) \to r)$.

**Proposition 12 (Cabalar et al., 2009)** *If $\varphi$ is function free, safe, and prenex formula, then: $\langle(D, I), T, T\rangle$ is an equilibrium model of $\varphi$ if and only if it is an equilibrium model of $\mathrm{Gr}_C(\varphi)$ (the grounding of $\varphi$ over $C$).*

## 5.2 Interpolation

On the basis of Proposition 12 we could already establish interpolation theorems for safe formulas in prenex form, essentially by replacing such formulas by their ground versions and working in propositional logic. However, we can also apply Propositions 9 and 10 directly by noting the property shown by Cabalar et al. (2009) that safe prenex formulas have definable classes of equilibrium models.

**Theorem 3 (interpolation for safe formulas)** *Safe formulas in prenex form have $\mathbf{QHT}^s_=$-definable classes of equilibrium models. Therefore for such formulas $(\mathord{\sim}, \mathord{\sim})$-interpolation for $\mathord{\sim}_{cw}$ inference holds as in Proposition 9 and $(\mathord{\sim}, \vdash)$-interpolation holds for $\mathord{\sim}_{ow}$ inference as in Proposition 10.*

## 6. Interpolation in Answer Set Semantics

Answer set programming (ASP) has become an established form of declarative, logic-based programming and its basic ideas are now well-known. For a textbook treatment the reader is referred to Baral's book (2003). As is also well-known, the origins of ASP lie in the stable model and answer set semantics for logic programs introduced by Gelfond and Lifschitz (1988, 1990, 1991). This semantics made use of a fixpoint condition involving a certain 'reduct' operator. Subsequent extensions of the concept to cover more general kinds of rules also relied on a reduct operator of similar sort. For the original definitions, the reader is referred to the various papers cited.

---

8. Recall that a subexpression of a formula is said to be positive in it if the number of implications that contain that subexpression in the antecedent is even, and negative if it is odd. Here we also consider that $\neg\varphi$ is defined as $\varphi \to \bot$.

The correspondence between answer set semantics and equilibrium logic is also well-established and has been discussed in many publications, beginning with Pearce (1997), who first showed how the answer sets of disjunctive programs can be regarded as equilibrium models (Lifschitz et al., 2001, 2007; Ferraris et al., 2007; Pearce & Valverde, 2005, 2006, 2008). For our purposes it will suffice to recall just two important syntactic classes of programs and the main features of the correspondence with equilibrium logic.

At one extreme we have ground, disjunctive logic programs; we treat them here without strong negation, so their answer sets are simply collections of atoms. These programs consist of sets of ground rules of the form

$$K_1 \vee \ldots \vee K_k \leftarrow L_1, \ldots L_m, not L_{m+1}, \ldots, not L_n \tag{7}$$

where the $L_i$ and $K_j$ are atoms. The 'translation' from the syntax of programs to **HT** propositional formulas is the trivial one, viz. (7) corresponds to the **HT** sentence

$$L_1 \wedge \ldots \wedge L_m \wedge \neg L_{m+1} \wedge \ldots \wedge \neg L_n \rightarrow K_1 \vee \ldots \vee K_k \tag{8}$$

Under this translation the correspondence between the answer sets and the equilibrium models of ground disjunctive programs is also the direct one:

**Proposition 13** *Let $\Pi$ be a disjunctive logic program. Then $\langle T, T \rangle$ is an equilibrium model of $\Pi$ if and only if $T$ is an answer set of $\Pi$.*

This was first shown by Pearce (1997) but the basic equivalence was later shown to hold for more general classes of programs by Pearce, P. de Guzman and Valverde (2000).

It is also common to treat non-ground rules of form (7) where variables may appear. These variables are thought of as being universally quantified, so the corresponding translation into a logical formula would simply be the universal closure of formula (8).

At the other extreme, Ferraris, Lee and Lifschitz (2007) provided a new definition of stable model for arbitrary first-order formulas. In this case the property of being a stable model is defined syntactically via a second-order condition that resembles parallel circumscription. However they also showed that the new notion of stable model is equivalent to that of equilibrium model as defined here for first-order languages. In a sequel to this paper, Lee, Lifschitz and Palla (2008a) have applied the new definition and made the following refinements. The *stable models* of a formula are defined as Ferraris et al. (2007) were, while the *answer sets* of a formula are those Herbrand models of the formula that are stable in the sense of Ferraris et al. Using this new terminology, it follows that in general stable models and equilibrium models coincide, while answer sets are equivalent to SNA-**QHT** models that are equilibrium models.

In between these two extremes many syntactically different kinds of programs have been considered and several variations in the concept of answer set have been proposed. However all the main varieties display a similar correspondence to equilibrium logic. It is merely necessary in some cases to restrict attention to specific kinds of equilibrium models, e.g. Herbrand models, UNA-models or SNA-models. It is important to notice also that this correspondence extends to many of the additional constructs that have been introduced in ASP, such as cardinality and weight constraints and even general forms of aggregates (Lee & Meng, 2009). All these can be accommodated in equilibrium logic by translation into logical formulas.

In ASP the main emphasis is on finding answer sets and this is what most answer set solvers compute. Less attention is placed on implementing a non-monotonic inference relation or a query answering mechanism. However there is a standard, skeptical concept of inference or entailment associated with answer set semantics. This notion of entailment or consequence for programs under the answer set semantics is that a query $Q$ is entailed by a program $\Pi$ if $Q$ is true in all answer sets of $\Pi$ (Balduccini, Gelfond, & Nogueira, 2000). Let us denote this entailment or consequence relation by $\vdash_{AS}$. Evidently atoms are true in an answer set if and only if they belong to it. Conjunctions and disjunctions are handled in the obvious way (Lifschitz, Tang, & Turner, 1999; Balduccini et al., 2000). Sometimes, queries of the form *not a*, or in logical notation $\neg a$, are not explicitly dealt with (Balduccini et al., 2000). However it seems to be in keeping with the semantics to regard a formula of form *not α* or $\neg \alpha$ to be true in an answer set if and only if $\alpha$ is not true. Another way to express this would be to say that an answer set satisfies $\neg \alpha$ if it does not violate the constraint $\{\leftarrow \alpha\}$, understanding constraint violation as Lifschitz, Tang and Turner (1999).[9] In this way we would say that $\Pi \vdash_{AS} \neg A$ if no answer set of $\Pi$ contains $A$. Similarly, the interpretation of queries containing quantifiers in answer set semantics should also conform to that of equilibrium logic, taking account of any specific restrictions, such as Herbrand models, that might be imposed.

We can therefore transfer interpolation properties from equilibrium logic to answer set semantics and ASP. It remains to consider whether $\vdash_{AS}$ is best identified with the closed world version of inference, $\vdash_{cw}$, or the more open world version, $\vdash_{ow}$. Again, since ASP solvers do not generally implement inference engines, the difference is largely a theoretical one. In traditional logic programming, however, a query that does not belong to the language of the program is usually answered *false*. It also seems quite natural in an ASP context that, given a program $\Pi$ and a query $Q$, one should consider the stable models of $\Pi$ in the language $\mathcal{L}(\Pi) \cup \mathcal{L}(Q)$ even if this is a proper extension of the language of $\Pi$.[10] So in general $\vdash_{cw}$ seems a natural choice for answer set inference. On the other hand, there are contexts where answer set semantics is used in a more open world setting, for example in the setting of hybrid knowledge bases (Rosati, 2005) where non-monotonic rules are combined with ontologies formalised in description logics. For such systems a semantics in terms of equilibrium logic was provided by de Bruijn, Pearce, Polleres and Valverde (2007). Here an entailment relation in the style of $\vdash_{ow}$ might sometimes be more appropriate.

In general answer set semantics is defined only for coherent programs or theories. For these, by identifying $\vdash_{AS}$ with $\vdash_{cw}$, we can apply Proposition 9 directly:

**Corollary 2** *For coherent formulas $\alpha$, $(\vdash, \vdash)$-interpolation in the form of Proposition 9 holds for entailment $\vdash_{AS}$ in answer set semantics.*

---

9. In logical terms this constraint would be written $\alpha \to \bot$.

10. Notice that by Proposition 12 if a program consists of safe formulas, an atomic query $q(a)$ is automatically false if $a$ does not belong to the language of the program (even if $q$ does), simply because grounding with the program constants is sufficient to generate all answer sets.

## 7. An Application of Interpolation

The Interpolation property has been applied in various areas of computer science, notably in software specification (Bicarregui et al., 2001) and in the construction of formal ontologies (Lutz & Wolter, 2010). In both areas it is relevant to modularity issues. Here we discuss a simple application related to a concept described by Lutz and Wolter that we can adapt to the case of nonmonotonic logic programs.

One way to compare two theories is via their nonmonotonic consequence relations. When two theories produce the same answers for a given query language, we can call them *inseparable*; this term is used in mathematical logic and also in the study of formal ontologies (Lutz & Wolter, 2010).

Let us say therefore that $\Pi_1$ and $\Pi_2$ are $\mathcal{L}$-inseparable if for any $\varphi$ such that $V(\varphi) \subseteq \mathcal{L}$, $\Pi_1 \mathrel{\vert\!\sim} \varphi \Leftrightarrow \Pi_2 \mathrel{\vert\!\sim} \varphi$.

**Proposition 14** *Let $\Pi_1$ and $\Pi_2$ be $\mathcal{L}$-inseparable theories such that $V(\Pi_1) = V(\Pi_2) = V$, say. Then for any $\mathcal{L}' \supset \mathcal{L}$ such that $V \cap \mathcal{L}' \subseteq \mathcal{L}$, $\Pi_1$ and $\Pi_2$ are $\mathcal{L}'$-inseparable.*

*Proof.* Assume that $\Pi_1$ and $\Pi_2$ are $\mathcal{L}$-inseparable and that $\mathcal{L}'$ is an extension of $\mathcal{L}$ such that $V \cap \mathcal{L}' \subseteq \mathcal{L}$. Suppose $\Pi_1 \mathrel{\vert\!\sim} \varphi$, where $V(\varphi) = \mathcal{L}'$. Suppose $\mathcal{L}' \setminus V = \{B_1, \ldots B_n\}$. By Proposition 7 there is an interpolant $\gamma$ for $(\Pi_1, \varphi)$ such that $\gamma \models \neg B_1 \wedge \ldots \wedge \neg B_n \to \varphi$. Since $\Pi_1 \mathrel{\vert\!\sim} \gamma$ and $V(\gamma) \subseteq \mathcal{L}$, by $\mathcal{L}$-inseparability we have $\Pi_2 \mathrel{\vert\!\sim} \gamma$. By right absorption therefore $\Pi_2 \mathrel{\vert\!\sim} \neg B_1 \wedge \ldots \wedge \neg B_n \to \varphi$. However it is clear that $B_1, \ldots B_n$ are false in all equilibrium models of $\Pi_2$, so $\Pi_2 \mathrel{\vert\!\sim} \varphi$. Repeating this argument with $\Pi_1$ and $\Pi_2$ interchanged shows that the theories are $\mathcal{L}'$-inseparable. $\qquad\square$

The above proof is similar to the argument given by Lutz and Wolter (2010) for Theorem 7 of that paper, applied to TBoxes in description logics. The property described is called there *robustness under signature extensions*. Notice however that, since $\mathrel{\vert\!\sim}$ is not in general transitive we cannot immediately infer from $\Pi_2 \mathrel{\vert\!\sim} \gamma$ and $\gamma \mathrel{\vert\!\sim} \varphi$ that also $\Pi_2 \mathrel{\vert\!\sim} \varphi$. This highlights the added strength of using explicitly the set $\{B_1, \ldots B_n\}$ and the property that **HT** forms a deductive basis for $\mathrel{\vert\!\sim}$.

In the study of modularity and logical relations between programs in ASP, it is more common to compare their sets of answer sets rather than their consequence classes. However it turns out that the notion of inseparability is very close to a concept that has already been studied in ASP. Two theories or programs are said to be projectively equivalent if the projections of their answer sets onto a common sublanguage agree (Eiter, Tompits, & Woltran, 2005). Formally, let $\Pi_1, \Pi_2$ be theories and $\mathcal{L}$ be a signature such that $\mathcal{L} \subseteq V(\Pi_1) \cap V(\Pi_2)$. Then $\Pi_1$ and $\Pi_2$ are said to be *projectively equivalent relative to $\mathcal{L}$* if $E(\Pi_1){\upharpoonright}_{\mathcal{L}} = E(\Pi_2){\upharpoonright}_{\mathcal{L}}$, where for any class of models $\mathcal{K}$, $\mathcal{K}{\upharpoonright}_{\mathcal{L}} = \{\mathcal{M}{\upharpoonright}_{\mathcal{L}} : \mathcal{M} \in \mathcal{K}\}$.

**Proposition 15** *Let $\Pi_1, \Pi_2$ be theories and $\mathcal{L}$ a signature such that $\mathcal{L} \subseteq V(\Pi_1) \cap V(\Pi_2)$. $\Pi_1$ and $\Pi_2$ are projectively equivalent relative to $\mathcal{L}$ if and only if they are $\mathcal{L}$-inseparable.*

In other words these two concepts agree whenever $\mathcal{L}$ is a common sublanguage of $\Pi_1, \Pi_2$. The main advantage of $\mathcal{L}$-inseparability is that it seems the more natural one to use if we want to consider signatures that extend the language of either program or theory.

## 8. Uniform Interpolation and Forgetting

A stronger form of interpolation known as *uniform* interpolation is also important for certain applications in computer science (Konev et al., 2009). As usual, given $\alpha, \beta$ with $\alpha \vdash \beta$, we are interested in interpolants $\gamma$ such that

$$\alpha \vdash \gamma \quad \& \quad \gamma \vdash \beta \tag{9}$$

where $\gamma$ contains only predicate and constant symbols that belong to both $\alpha$ and $\beta$. The difference now is that $\gamma$ is said to be a *uniform interpolant* if (9) holds *for any* $\beta$ in the same signature such that $\alpha \vdash \beta$. A logic is said to have the uniform interpolation property if such uniform interpolants exist for all $\alpha, \beta$.

In classical propositional logic, the uniform interpolation holds, however it fails in first order classical logic and in many non-classical logics. It may hold when certain restrictions are placed on the theory language in which $\alpha$ is formulated and on the query language containing $\beta$. For example it has been shown to hold for some description logics (Kontchakov et al., 2008) where such syntactic restrictions apply. Even in ASP it turns out that a form of uniform interpolation holds for a very restricted query language, essentially one that allows just instance retrieval. We can show this by using some known results in ASP about the concept of *forgetting* (Eiter & Wang, 2008) that is quite closely related to interpolation.

Variable forgetting, as studied by Eiter and Wang (2008), is concerned with the following problem. Given a disjunctive logic program $\Pi$ and a certain atom $a$ occurring in $\Pi$, construct a new program, to be denoted by **forget**$(\Pi, a)$, that does not contain $a$ but whose answer sets are in other respects as close as possible to those of $\Pi$. For the precise notion of closeness the reader is referred to paper of Eiter and Wang, however some consequences will be evident shortly. Eiter and Wang define **forget**$(\Pi, a)$ (as a generic term), show that such programs exist whenever $\Pi$ is coherent, and provide different algorithms to compute such programs.

Given coherent $\Pi$ and $a$ in $\Pi$, the results **forget**$(\Pi, a)$, of forgetting about $a$ in $\Pi$ may be different but are always answer set equivalent. Moreover for our purposes they satisfy the following key property, where $\Pi$ is coherent, $a, b$ are distinct atoms in $\Pi$ and as usual $\vdash\!\!\sim$ denotes nonmonotonic consequence,

$$\Pi \vdash\!\!\sim b \quad \Leftrightarrow \quad \textbf{forget}(\Pi, a) \vdash\!\!\sim b. \tag{10}$$

showing that indeed the answer sets of $\Pi$ and **forget**$(\Pi, a)$ are closely related.

To establish a version of uniform interpolation for the case of disjunctive programs and simple, atomic queries, we need to show that we can always find a $\Pi' = \textbf{forget}(\Pi, a)$ such that $\Pi \vdash\!\!\sim \Pi'$. For this we can examine the first algorithm of Eiter and Wang for computing **forget**$(\Pi, a)$; this is also the simplest of the three algorithms presented. Let $\Pi$ be a coherent program with rules of form (7) that we write as formulas of form (8) and let $a$ be an atom in $\Pi$. The method for constructing a $\Pi' = \textbf{forget}(\Pi, a)$ is as follows.

1. Compute the equilibrium models $E(\Pi)$.

2. Let $E'$ be the result of removing $a$ from each $\mathcal{M} \in E(\Pi)$.

3. Remove from $E'$ any model that is non-minimal to form $E''(= \{A_1, \ldots, A_m\}$, say).

4. Construct a program $\Pi'$ whose answer sets are precisely $\{A_1, \ldots, A_m\}$ as follows:

- for each $A_i$, set $\Pi_i = \{\neg \overline{A_i} \rightarrow a' : a' \in A_i\}$, where $\overline{A_i} = V(\Pi) \setminus A_i$.
- Set $\Pi' = \Pi_1 \cup \ldots \cup \Pi_m$.

We can now verify the desired property. Let $\mathcal{L}$ be the simple query language composed of conjunctions of literals.

**Proposition 16** *In equilibrium logic (or answer set programming) uniform interpolation holds for (coherent) disjunctive programs and queries in $\mathcal{L}(V(\Pi))$.*

*Proof.* To prove the claim we shall show the following. Let $\Pi$ be a coherent disjunctive program and let $\mathcal{L} = \mathcal{L}(V)$ for some $V \subseteq V(\Pi)$. Then there is a program $\Pi'$ such that $V(\Pi') = V$ and for any $\varphi \in \mathcal{L}$,

$$\Pi \mathrel{|\!\sim} \varphi \Rightarrow (\Pi \mathrel{|\!\sim} \Pi' \ \& \ \Pi' \mathrel{|\!\sim} \varphi)$$

To begin, let $\Pi$ and $\varphi$ be as above with $\Pi \mathrel{|\!\sim} \varphi$. Let $X = \{a_1, \ldots, a_n\} = V(\Pi) \setminus V$. Then we choose $\Pi'$ to be the result of forgetting about $X$ in $\Pi$, defined by Eiter and Wang (2008) as follows:

$$\mathbf{forget}(\Pi, X) := \mathbf{forget}(\mathbf{forget}(\mathbf{forget}(\Pi, a_1), a_2), \ldots, a_n),$$

and it is shown there that the order of the atoms in $X$ does not matter. Now we know by (10) that for any atom $a \in V$ and any $i = 1, n$,

$$\Pi \mathrel{|\!\sim} a \Leftrightarrow \mathbf{forget}(\Pi, a_i) \mathrel{|\!\sim} a, \tag{11}$$

therefore

$$\Pi \mathrel{|\!\sim} a \Rightarrow \mathbf{forget}(\Pi, X) \mathrel{|\!\sim} a. \tag{12}$$

Let $\Pi'$ be $\mathbf{forget}(\Pi, X)$ as determined by algorithm 1 of Eiter and Wang (2008) described above. It is easy to see by (11) and the semantics of $\mathrel{|\!\sim}$ that (12) continues to hold where $a$ is replaced by a negated atom $\neg b$ and therefore also by any conjunction of literals since a conjunction is entailed only if each element holds in every equilibrium model.[11] So it remains to show that $\Pi \mathrel{|\!\sim} \Pi'$. Again, it will suffice to show this entailment for one member of the sequence $\mathbf{forget}(\Pi, a_i)$ and since the order is irrelevant wlog we can choose the first element $\mathbf{forget}(\Pi, a_1)$ and show that $\Pi \mathrel{|\!\sim} \mathbf{forget}(\Pi, a_1)$. We compute the programs $\Pi_1, \ldots, \Pi_m$ as in the algorithm. Then we need to check that $\Pi \mathrel{|\!\sim} \Pi_i$ for any $i = 1, n$, i.e. that for each $\mathcal{M} \in E(\Pi)$, $\mathcal{M} \models \{\neg \overline{A_i} \rightarrow a' : a' \in A_i\}$.

Consider $\mathcal{M} \in E(\Pi)$ where $\mathcal{M} = \langle T, T \rangle$. We distinguish two cases. (i) $A_i \subseteq T$. Then $\mathcal{M} \models a'$ for each $a' \in A_i$. It follows that $\mathcal{M} \models \neg \overline{A_i} \rightarrow a'$ for each $a' \in A_i$ and so $\mathcal{M} \models \{\neg \overline{A_i} \rightarrow a' : a' \in A_i\}$. Case (ii) $A_i \not\subseteq T$. Then $T$ and $A_i$ are incomparable. In particular we cannot have $T \subset A_i$ by the minimality property of $A_i$ obtained in step 3. Hence $T \cap \overline{A_i} \neq \varnothing$. Choose $a'' \in T \cap \overline{A_i}$. Then $\mathcal{M} \models a''$, so $\mathcal{M} \not\models \neg a''$ and hence $\mathcal{M} \not\models \neg \overline{A_i}$. Consequently, for any $a'$, $\mathcal{M} \models \neg \overline{A_i} \rightarrow a'$ and so $\mathcal{M} \models \{\neg \overline{A_i} \rightarrow a' : a' \in A_i\}$. It follows that for any $i$, $\Pi \mathrel{|\!\sim} \Pi_i$ and so by construction $\Pi \mathrel{|\!\sim} \Pi'$, which establishes the proposition. $\square$

---

11. As Eiter and Wang (2008) point out, if an atom $b$ is true in some answer set of $\mathbf{forget}(\Pi, a)$, then it must also be true in some answer set of $\Pi$, showing that (12) holds for literals.

## 8.1 Extending the Query Language

If we establish uniform interpolation in ASP using the method of forgetting, as defined by Eiter and Wang (2008), it seems clear that we cannot extend in a non-trivial way the expressive power of the query language $\mathcal{L}$. Since the method of forgetting $a$ in $\Pi$ removes non-minimal sets from $E(\Pi)$ (once $a$ has been removed), an atom $b$ might be true in some equilibrium model of $\Pi$ but not in any equilibrium model of $\mathbf{forget}(\Pi, a)$. Hence we might have a disjunction, say $a \vee b$, derivable from $\Pi$ but not from $\mathbf{forget}(\Pi, a)$. Likewise, if we consider programs with variables in a first-order setting, we cannot in general extend $\mathcal{L}$ to include existential queries.

On the other hand, the property of uniform interpolation certainly holds for any $\mathcal{L}(V)$ even without the condition that $V \subseteq V(\Pi)$. Suppose that $\Pi \hspace{1pt}\vert\hspace{-3pt}\sim \varphi$ where $V(\varphi) \setminus V(\Pi) \neq \varnothing$, say $V(\varphi) \setminus V(\Pi) = \{b_1, \ldots, b_k\}$. Then $b_1, \ldots, b_k$ are false in all equilibrium models of $\Pi$. Trivially, if $b$ is not in $V(\Pi)$ we can regard the result of forgetting about $b$ in $\Pi$ as just $\Pi$. So we can repeat the proof of Proposition 16, but now setting $X = \{V(\Pi) \setminus V\} \cup \{V \setminus V(\Pi)\}$. All the relevant properties will continue to hold.

An interesting open question is whether we can extend the theory language to include more general kinds of program rules such as those allowing negation in the head. Accommodating these kinds of formulas would constitute an important generalisation since they amount to a normal form in equilibrium logic. However, the answer sets of such programs do not satisfy the minimality property that holds for the answer sets of disjunctive programs, so it is clear that the definition of forgetting would need to be appropriately modified - a task that we do not attempt here.

## 9. Literature and Related Work

The interpolation theorem for classical logic is due to Craig (1957); it was extended to intuitionistic logic by Schütte (1962). Maksimova (1977) characterised the super-intuitionistic propositional logics possessing interpolation. A modern, comprehensive treatment of interpolation in modal and intuitionistic logics can be found in the monograph of Gabbay and Maksimova (2005).

In non-monotonic logics, interpolation has received little attention. A notable exception is an article (Amir, 2002) establishing some interpolation properties for circumscription and default logic. By the well-known relation between the answer sets of disjunctive programs and the extensions of corresponding default theories, he also derives a form of interpolation for ASP. With regard to answer set semantics, the approach of Amir is quite different from ours. Since it is founded on an analysis of default logic, it uses classical logic as an underlying base. So Amir's version of interpolation is a form of (3) where $L$ is classical logic; there is no requirement that $\vdash_L$ form a well-behaved sublogic of $\hspace{1pt}\vert\hspace{-3pt}\sim$, e.g. a deductive base. As Amir remarks, one cannot deduce in general from property (4) that $\alpha \hspace{1pt}\vert\hspace{-3pt}\sim \beta$. However if $L$ is classical logic one cannot even deduce $\alpha \hspace{1pt}\vert\hspace{-3pt}\sim \beta$ from (3). More generally, there is no counterpart to our Proposition 1 in this case. Another difference with respect to our approach is that Amir does not discuss the nature of the $\hspace{1pt}\vert\hspace{-3pt}\sim$ relation for ASP in detail, in particular how to understand $\Pi \hspace{1pt}\vert\hspace{-3pt}\sim \varphi$ in case $\varphi$ contains atoms not present in the program $\Pi$. In fact, if we interpret $\hspace{1pt}\vert\hspace{-3pt}\sim_{AS}$ as in Section 6 above, it is easy to refute $(\hspace{1pt}\vert\hspace{-3pt}\sim, \vdash_L)$-interpolation where $L$ is classical logic. Let $\Pi$ be the program $B \leftarrow \neg A$ and $q$ the query $B \wedge \neg C$. Then clearly $\Pi \hspace{0.1em}\vert\hspace{-0.3em}\sim_{AS} q$, but there is no formula in the vocabulary $B$ that would classically entail $\neg C$. Under any interpretation of answer set inference such that atoms not in the program are regarded as false, $(\hspace{0.1em}\vert\hspace{-0.3em}\sim, \vdash_L)$-interpolation would be refuted.

## 10. Conclusions

We have discussed two kinds of interpolation properties for non-monotonic inference relations and shown that these properties hold in turn for the two different inference relations that we can associate with propositional equilibrium logic. In each case we use the fact that the collection of equilibrium models is definable in the logic **HT** of here-and-there and that this logic possesses the usual form of interpolation. One of the forms of inference studied seems to be in many cases an appropriate concept to associate with answer set programming, although in general ASP systems are not tailored to query answering or deduction. Using results of Eiter and Wang (2008) about variable forgetting in ASP, we could also show how the property of uniform interpolation holds for disjunctive programs and a restricted query language.

We have also discussed the interpolation property for first-order equilibrium logic based on a quantified version **QHT** of the logic of here-and-there, obtaining analogous results as for the propositional case whenever the collection of equilibrium models is definable. These positive results transfer to answer set programming under the assumption usually made in ASP systems that programs are *safe* and therefore have definable collections of answer sets. As we saw, the notion of safety can be quite generally defined for theories and is not limited to normal or disjunctive programs.

### Acknowledgments

David Pearce is partially supported by MEC projects TIN2009-14562-C05-02 and CSD2007-00022. Agustín Valverde is partially supported by MEC project TIN2009-14562-C05-01, and Junta de Andalucia projects P09-FQM-05233 and TIC-115. The authors are grateful to the anonymous referees for helpful comments.